\documentstyle[preprint,aps,graphicx,floats]{revtex}

\begin{document}

\preprint{$
\begin{array}{l}
\mbox{UB-HET-04-01}\\[-3mm]
\mbox{NSF-KITP-04-45}\\[-3mm]
\mbox{May~2004} \\ [8mm]
\end{array}
$}

\title{Electroweak Radiative Corrections to
$p\,p\hskip-7pt\hbox{$^{^{(\!-\!)}}$} \to W^\pm\to\ell^\pm\nu$\\ Beyond
the Pole Approximation\\[1cm]}

\author{U.~Baur\footnote{e-mail:
baur@ubhex.physics.buffalo.edu} and
D.~Wackeroth\footnote{e-mail: dow@ubpheno.physics.buffalo.edu}\\[3.mm]} 
\address{Department of Physics,
State University of New York at Buffalo,\\ Buffalo, NY 14260, USA\\[1.5cm]}

\maketitle 

\begin{abstract}
\baselineskip15.pt  %to keep abstract on 1 page
We present a calculation of the complete electroweak ${\cal O}(\alpha)$
corrections to 
$p\,p\hskip-7pt\hbox{$^{^{(\!-\!)}}$} \to W^\pm\to\ell^\pm\nu X$
($\ell=e,\,\mu$) in the Standard Model of electroweak interactions,
focusing on those corrections which do not contribute in the pole
approximation. We
study in detail the effect of these corrections on the
transverse mass distribution, the $W$-width measurement, and the 
transverse mass ratio and cross section ratio of $W$ and $Z$ bosons.
\end{abstract}

\newpage

%%%%%%%%%%%%%%%%%%%%%%%%%%%%%%%  MAIN TEXT  %%%%%%%%%%%%%%%%%%%%%%%%%%%%

\tightenlines

\section{Introduction}
\label{sec:sec1}

The Standard Model of electroweak interactions (SM) so far withstood
all experimental challenges and is tested as a quantum field theory at
the 0.1\% level~\cite{lepewwg}. However, the mechanism of mass
generation in the SM predicts the existence of a Higgs boson which, so
far, has eluded direct observation. Direct searches at LEP2 give 
a 95\% confidence-level lower bound on the mass of the
SM Higgs boson of $M_H> 114.4$~GeV~\cite{Barate:2003sz}. Indirect
information on the mass of the Higgs boson can be extracted from the
$M_H$ dependence of radiative corrections to the W boson mass. With
the present knowledge of the $W$ boson and top quark
masses~\cite{lepewwg,mtop}, and the electromagnetic coupling constant,
$\alpha(M_Z^2)$~\cite{Eidelman:1995ny}, the SM 
Higgs boson mass can be indirectly constrained to 
$M_H=113^{+62}_{-42}$~GeV~\cite{lepewwg,clare} by a global fit to all 
electroweak precision
data.  Future more precise measurements of the $W$ boson and top quark
masses are expected to  considerably improve the present indirect bound on
$M_H$: with a precision of 27~MeV for the $W$ boson mass, $M_W$, and
2.7~GeV for the top quark mass, which are target values for the expected 
integrated luminosity of 2~fb$^{-1}$ of 
Run~II of the Tevatron, $M_H$ can be predicted with an
uncertainty of about $35\%$~\cite{blueband}. In addition, the
confrontation of a precisely measured $W$ boson mass with the
indirect SM prediction from a global fit to all 
electroweak precision data, $M_W=80.386 \pm 0.023$~GeV~\cite{lepewwg} 
will provide a stringent test of the SM. A precise measurement of
the $W$ width, $\Gamma_W$, and comparison with the SM prediction, will
help to further scrutinize the SM. The $W$ width is expected to be
measured in Run~II with a precision of about $25-30$~MeV from the high 
transverse mass tail when data from both lepton channels and both
experiments are combined~\cite{Brock:1999ep}.
It can also be determined indirectly from the 
cross section ratio~\cite{D0Wcross,cdfr},
\begin{equation}
R_{W/Z}={\sigma(p\bar p\to W\to\ell\nu X)\over\sigma(p\bar p\to Z
\to\ell^+\ell^- X)}~,
\label{eq:ratio}
\end{equation}
together with the theoretical prediction for the ratio of the total $W$
and $Z$ production cross sections, the LEP measurement of the
branching ratio $B(Z\to\ell^+\ell^-)$~\cite{lepewwg}, and the SM
prediction for the $W\to\ell\nu$ decay width. 

In hadronic collisions, the $W$ mass is usually determined from the
transverse mass 
distribution of the lepton and neutrino which originate from the $W$
decay, $W\to\ell\nu$. The experimental
uncertainty on $M_W$ strongly depends on an accurate measurement of the
transverse momentum of the neutrino which requires that
the transverse momentum distribution of the $W$ is well understood. In
lowest order, the $W$ is produced without any transverse momentum. Only
when QCD corrections are taken into account does the $W$ acquire a
non-zero transverse momentum, $p_T^W$. For a detailed understanding of
the $p_T^W$ distribution, it is necessary to resum the soft gluon
emission terms~\cite{resbos}, and to model non-perturbative QCD
corrections~\cite{brock}. In
addition to QCD corrections, electroweak (EW) radiative corrections play an
important role in the $W$ mass and width measurement; final state photon
radiation is known to shift both quantities by ${\cal
O}(100$~MeV)~\cite{cdfwmass,d0wmass,unknown:2003sv,Abe:1994qn,Affolder:2000mt,Abazov:2002xj}. 
In order to measure the $W$ mass and width with high
precision at a hadron collider, it is thus necessary to fully 
understand and control higher order QCD and electroweak 
corrections. In the last few years, significant progress in our
understanding of the EW corrections to $W$ boson production in
hadronic collisions has been made.
A calculation of the ${\cal O}(\alpha)$ EW corrections to 
$p\,p\hskip-7pt\hbox{$^{^{(\!-\!)}}$} \to W^{\pm} \to\ell^{\pm} \nu$
($\ell=e,\,\mu$) in the pole approximation was presented in
Ref.~\cite{Baur:1999kt}. The complete ${\cal O}(\alpha)$ corrections 
were calculated in Ref.~\cite{Dittmaier:2001ay}. Two photon radiation in
$W$ production and decay has been calculated in
Ref.~\cite{Baur:1999hm}. Finally, first steps towards
going beyond fixed order in EW radiative corrections were taken in
Refs.~\cite{CarloniCalame:2003ux} and~\cite{Placzek:2003zg}
by including the effects of final state multiphoton radiation in $W$
production, and in Ref.~\cite{Cao:2004yy} where final state photon
radiation was added to a calculation of $W$ boson production which 
includes resummed QCD corrections. 

In this paper we
present an independent calculation of the complete ${\cal O}(\alpha)$ EW
radiative corrections to $p\,p\hskip-7pt\hbox{$^{^{(\!-\!)}}$}\to
W^{\pm} \to\ell^{\pm} \nu$ using the methods developed in
Ref.~\cite{Wackeroth:1996hz}, and examine how the non-resonant 
corrections neglected in Ref.~\cite{Baur:1999kt} affect several
observables of interest, in particular the measurement of the width of
the $W$ boson. Preliminary results of our calculation were reported in
Ref.~\cite{Baur:2002fn}. For the numerical evaluation, 
we use the Monte Carlo phase space slicing method for next-to-leading-order 
(NLO) calculations described in Ref.~\cite{NLOMC}. With the Monte Carlo 
method, it is easy to calculate a variety of observables simultaneously 
and to simulate detector response. The collinear 
singularities associated with initial state photon radiation are removed
by universal collinear counter terms
generated by ``renormalizing'' the parton distribution
functions (PDFs)~\cite{Baur:1999kt,rujula,perlt,qedhs,roth}, in complete
analogy to gluon emission in QCD. Final state
charged lepton mass effects are included in our calculation in the 
following approximation. The lepton mass regularizes the collinear 
singularity associated with final state photon radiation. The associated
mass singular logarithms of the form $\ln(\hat s/m_\ell^2)$, where $\hat
s$ is the squared parton center of mass energy and $m_\ell$ is the
charged lepton mass, are included in our calculation, but the very small
terms of ${\cal O}(m_\ell^2/\hat s)$ are neglected. 

The technical details of our calculation are described in Sec.~II. The
electroweak ${\cal O}(\alpha)$ corrections consist of the electroweak
one-loop 
contributions, including virtual photons, and of the emission of a real
photon. To regularize the ultraviolet divergences associated with the
virtual corrections, we use dimensional regularization in the {\sc
on-shell} renormalization scheme~\cite{bo86}. The non-resonant
corrections are small in the $W$ pole region, but 
become important at high $\ell\nu$ invariant masses, $m(\ell\nu)$, due
to the presence of large Sudakov-like electroweak
logarithms of the form $(\alpha/\pi)\ln^2(m(\ell\nu)/M_V)$ ($V=W,\,
Z$)~\cite{cc}. 

Numerical results for the Tevatron ($p\bar p$ collisions at
$\sqrt{s}=2$~TeV) and the CERN Large Hadron collider (LHC,
$pp$ collisions at $\sqrt{s}=14$~TeV) are presented in Sec.~III. For
$\ell\nu$ transverse masses above the $W$ peak region, the non-resonant
corrections reduce the differential cross section by ${\cal
O}(10\%)$. We study in detail how these corrections affect the
measurement of the $W$ width from the tail of the transverse mass
distribution. Using the results of Ref.~\cite{Baur:2001ze}, we also
consider how the ratio of the $W$ and $Z$ cross 
sections, $R_{W/Z}$ (see Eq.~(\ref{eq:ratio})), and the transverse mass
ratio of $W$ and $Z$ bosons are influenced. Finally, our conclusions are
presented in Sec.~IV.

\section{The ${\cal O}(\alpha)$ Electroweak radiative corrections to $W$
production: Resonant and Non-resonant Contributions} 
\label{sec:sec2}

\begin{figure}[t!]
\vskip 16cm
\includegraphics{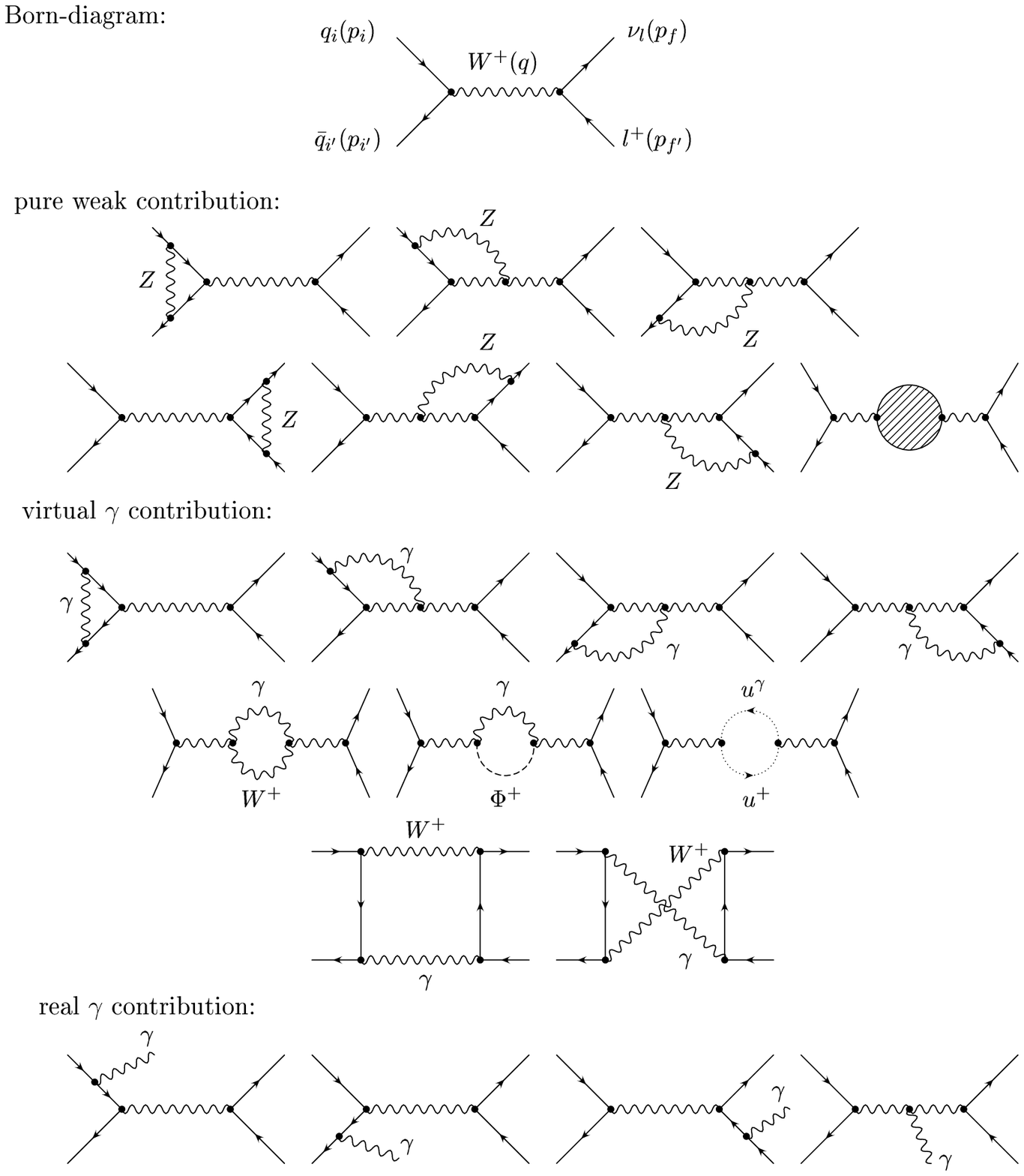}
\vspace*{1.cm}
\caption{The Feynman diagrams contributing to $W$ boson production at
${\cal O}(\alpha^3)$ in the Feynman - 't~Hooft gauge ($\Phi^+$: Higgs~--
ghost field, 
$u^+,u^{\gamma}$: Faddeev-Popov-ghost fields; the non-photonic contribution
to the $W$ self energy insertion is symbolized by the shaded loop).  
An explicit representation of the non-photonic contribution
to the $W$ self energy insertion can be found in
Ref.~[\ref{Hollik}].}
\label{fig:diag}
\end{figure}

\begin{figure}[t]
\begin{center}
\includegraphics[width=12cm]{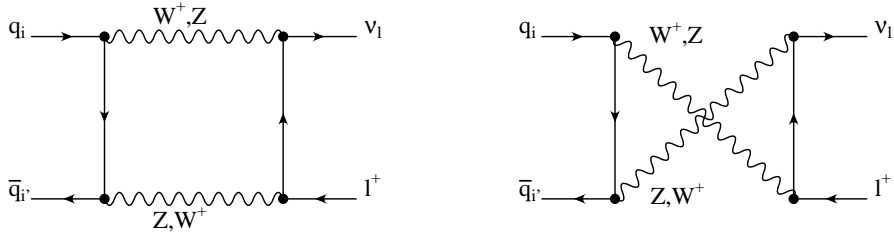}
%\vspace*{2mm}
\caption[]{\label{fig:wzbox} 
Feynman diagrams for the $W,Z$ box diagrams.}
\end{center}
\end{figure}

The complete ${\cal O}(\alpha^3)$ parton level cross section of $W$
production via the Drell-Yan mechanism $q_i
\overline{q}_{i'}\rightarrow f \bar f' (\gamma)$ is given by
\begin{eqnarray}\label{eq:full}
d \hat{\sigma}^{(0+1)}(\hat s,\hat t) & = &
d \hat \sigma^{(0)}+ d\hat \sigma_{virt} + 
\sum_{a=initial,final,\atop interf.} [d\hat\sigma^{(0)} \; F_{B\!R}^a\;+\; 
d \hat \sigma_{2\rightarrow 3}^a]\; ,
\end{eqnarray}
where the Born cross section, $d \hat \sigma^{(0)}$, is of
Breit-Wigner form and $\hat s$ and $\hat t$ are the usual Mandelstam
variables in the parton center of mass frame.  The corresponding
Feynman diagrams are shown in Fig.~\ref{fig:diag} and
Fig.~\ref{fig:wzbox}. 
$d\hat \sigma_{virt}$ describes the complete set
of virtual ${\cal O}(\alpha)$ contributions consisting of the pure
weak and the photonic virtual corrections:
\begin{equation}\label{eq:virt}
d \hat \sigma_{virt}(\hat s,\hat t)=d \hat \sigma^{(0)} 2 {\cal R}e  
[F_{weak}(\hat s) + F_{\gamma}(\hat s,\hat t)]+
 d \hat \sigma_{W\!Z box}(\hat s,\hat t) \; .
\end{equation}
Explicit expressions for the pure weak form factor $F_{weak}$ and for
the contribution of the $W,Z$ box diagrams, $d\hat \sigma_{W\!Z box}$,
are given in the Appendix. The form
factors $F_{B\!R}^a$ of Eq.~(\ref{eq:full}) describing the initial,
final state and interference contribution of real soft photon
radiation, and the photonic form factor $F_{\gamma}$
can be found in Ref.~\cite{Wackeroth:1996hz} (Appendix D).

The soft photon region is defined by requiring that the photon energy
in the parton center of mass frame, $\hat E_{\gamma}$, is smaller than a
cutoff $\Delta E=\delta_s\sqrt{\hat s}/2$.  In this phase space
region, the soft photon approximation can be used to calculate the
cross section as long as $\delta_s$ is sufficiently small.  Throughout
the calculation the soft singularities have been regularized by giving
the photon a fictitious mass. As usual, the
unphysical photon mass dependence cancels in the sum of the virtual and
soft photon terms, $2 {\cal R}e F_{\gamma} + \sum_a F_{B\!R}^a$.  The IR
finite contribution 
$d\hat\sigma_{2\rightarrow 3}^a$ describes real photon radiation with
$\hat E_{\gamma}>\Delta E$. The superscript $a$ denotes the initial state,
final state or interference contributions. Throughout the calculation
of the ${\cal O}(\alpha)$ corrections we consider
the fermions as being massless. We retain finite fermion masses only 
to regularize the collinear singularities which arise when
the photon is emitted collinear with one of the charged fermions.
Thus, $d\hat \sigma^{a}_{2\rightarrow 3}$ and the form factors 
$F_{B\!R}^{a}$ and $F_{\gamma}$
contain large mass singular logarithms which have to be
treated with special care. 
%For a detailed description of our treatment of these collinear
%singularities we refer to Ref.~\cite{Baur:1999kt}.
For final state photon radiation, the collinear singularity is
regularized by the finite lepton mass. In sufficiently inclusive
measurements, the mass singular logarithmic terms originating from the 
collinear singularity cancel~\cite{KLN}. For initial state photonic
corrections, however, the mass singular logarithms
always survive.  These singularities are universal to all orders in 
perturbation theory and can be absorbed by a redefinition ({\it
renormalization}) of the PDFs~\cite{Baur:1999kt,rujula}. This can be done in
complete analogy to the 
calculation of QCD radiative corrections.  As a result, 
the renormalized parton distribution functions become dependent on the QED
factorization scale $\mu_{\rm QED}$ which is controlled by the well-known
Gribov-Lipatov-Altarelli-Parisi (GLAP) equations \cite{ap}. These
universal photonic corrections can be taken into account by a
straightforward modification \cite{perlt,qedhs,roth} of the standard GLAP
evolution equations. 

The QED induced terms in the GLAP
equations lead to small corrections at the per-mille level 
to the distribution functions for most values of $x$ and
$\mu_{\rm QED}^2$~\cite{roth,Haywood:1999qg}. Only
at large $x \ {\stackrel{>}{\scriptstyle \sim}} \ 0.5$ and large
$\mu_{\rm QED}^2 \ {\stackrel{>}{\scriptstyle \sim}} \ 10^3$~GeV$^2$ do
the corrections reach the magnitude of one per cent. 

In order to treat the ${\cal O}(\alpha)$ initial state photonic corrections
to $W$ production in hadronic collisions in a consistent way, QED
corrections should be incorporated in the global fitting of the PDFs,
i.e. all data which are used to fit the parton distribution functions
should be corrected for QED effects. Current
fits~\cite{Martin:2001es,Pumplin:2002vw} to 
the PDFs do not include QED corrections. The missing QED corrections 
introduce an
uncertainty which, however, is probably much smaller than the present
experimental uncertainties on the parton distribution functions. 

Absorbing the collinear singularity into the PDFs introduces a
QED factorization scheme dependence. The squared matrix elements for different
QED factorization schemes differ by the finite ${\cal O}(\alpha)$ terms
which are absorbed into the PDFs in addition to the singular terms. Our
calculation has been carried out both in the 
QED ${\rm \overline{MS}}$ and DIS schemes, which are defined
analogously to the usual ${\rm \overline{MS}}$~\cite{MSBAR} and 
DIS~\cite{OWENSTUNG} schemes used in QCD calculations. All numerical
calculations in this paper are performed using the QED DIS scheme.
The QED DIS scheme is defined
by requiring the same expression for the leading and next-to-leading order 
structure function $F_2$ in deep inelastic scattering, which is given by
the sum of the quark distributions. Since $F_2$ data
are an important ingredient in extracting PDFs, the effect of
the ${\cal O}(\alpha)$ QED corrections on the PDFs should be reduced
in the QED DIS scheme.

In the vicinity of the $W$ resonance, the $W,Z$ box diagrams of
Fig.~\ref{fig:wzbox} can be neglected as non-resonant contributions of
higher order in perturbation theory, and, as demonstrated
in~\cite{Wackeroth:1996hz}, a gauge invariant decomposition of the
complete ${\cal O}(\alpha)$ contribution into a QED-like and a
modified weak part can be performed.  Unlike the $Z$ boson case, the
Feynman diagrams of Fig.~\ref{fig:diag} which involve a virtual photon
do not represent a gauge invariant subset.
In Ref.~\cite{Wackeroth:1996hz}, it was demonstrated that gauge invariant
contributions can be extracted from the infrared (IR) singular virtual
photonic corrections, $\tilde{F}_{Y\!F\!S}^a$ (the modified
Yennie-Frautschi-Suura form
factors).  These contributions can be combined with the 
real photon corrections in the soft photon region, $F_{B\!R}^a$, to
form gauge invariant QED-like contributions corresponding to initial
state, final state and interference corrections.  The IR finite
remainder of the virtual photonic corrections and the pure weak
one-loop corrections of Fig.~\ref{fig:diag} can be combined to
separately gauge invariant modified weak contributions to the $W$
boson production and decay processes.  

Both the QED-like and the
modified weak contributions can be expressed in terms of form factors,
$F_{Q\!E\!D}^{a}$ and $\tilde F_{weak}^{a}$, which multiply the Born
cross section~\cite{Wackeroth:1996hz}.  Thus, the complete ${\cal
O}(\alpha^3)$ parton level cross section of resonant $W$ production
via the Drell-Yan mechanism $q_i \overline{q}_{i'}\rightarrow f
f'(\gamma)$ can be expressed in the form
\begin{eqnarray}\label{eq:res}
d \hat{\sigma}^{(0+1)}_{res} &=& d \hat \sigma^{(0)}\; [1+ 2 {\cal R}e
(\tilde F_{weak}^{initial}+ 
\tilde F_{weak}^{final})(M_W^2)] 
\nonumber \\
&+ & \sum_{a=initial,final,\atop interf.} [d\hat\sigma^{(0)}\; 
F_{Q\!E\!D}^a(\hat s,\hat t)+
d \hat \sigma_{2\rightarrow 3}^a] \; .
\end{eqnarray}
The modified weak contributions have to be evaluated at $\hat
s=M_W^2$. Explicit expressions for the form
factors $F_{Q\!E\!D}^a,\tilde F_{weak}^a$ are given
in Ref.~\cite{Wackeroth:1996hz}.

\begin{figure}[t]
\begin{center}
\includegraphics[width=16cm]{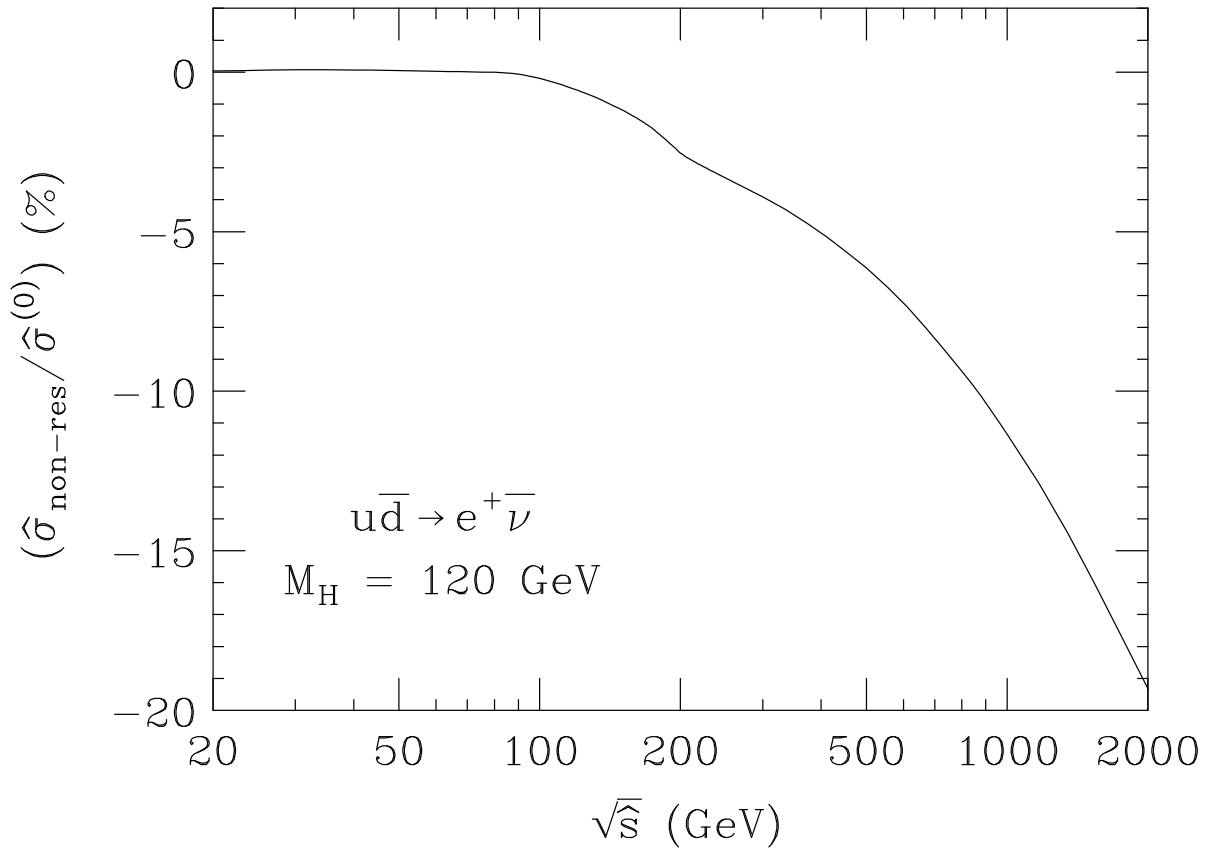}
\vspace*{2mm}
\caption[]{\label{fig:nonres} 
The relative size (in percent) of the non-resonant ${\cal
O}(\alpha)$ corrections to the Born $u \bar d \to W^+ \to  \ell^+\nu$
parton-level total cross section 
as a function of the parton center-of-mass energy, $\sqrt{\hat s}$. The
parameters used are listed in Eqs.~(\ref{eq:pars})~--~(\ref{eq:mhiggs}). }
\end{center}
\end{figure}

The non-resonant part, which was neglected 
in Refs.~\cite{Baur:1999kt} and~\cite{Wackeroth:1996hz}, can then be
obtained from the resonant contribution, $d\hat \sigma_{res}^{(0+1)}$ (see
Eq.~(\ref{eq:res})), and the complete ${\cal O}(\alpha)$ contribution,
$d\hat \sigma^{(0+1)}$ of Eq.~(\ref{eq:full}):
\begin{eqnarray}\label{eq:nonres}
d \hat{\sigma}_{non-res}(\hat s,\hat t) & = &d \hat{\sigma}^{(0+1)} - 
d \hat{\sigma}^{(0+1)}_{res}
\nonumber \\
&=& d \hat \sigma_{W\!Z box}(\hat s,\hat t)+ d \hat \sigma^{(0)} \,
2 {\cal R}e \left[ \sum_{a=initial,final} (F_{weak}^a(\hat s)-\tilde
F_{weak}^a(M_W^2))  
\right. \nonumber \\
&+& \left.
(F_{\gamma} - \sum_{a=initial,final,\atop interf.}
\tilde{F}^a_{Y\!F\!S})(\hat s,\hat t)\right] \; , 
\end{eqnarray}
where we have used Eq.~(\ref{eq:weakif}) and $F_{Q\!E\!D}^a=F^a_{B\!R}+2
{\cal R}e \tilde{F}^a_{Y\!F\!S}$ ($a=initial,final,interf.$).
Equation~(\ref{eq:nonres}) shows that the non-resonant contribution consists
of the $W,Z$ box contributions, the IR finite remnants of
the virtual photon one-loop corrections which are not included in the
YFS form factor, and the $\hat s$-dependent parts of the pure weak vertex and
self-energy one-loop corrections, which have been neglected when
evaluating the pure weak form factor at $\hat s=M_W^2$.  $d\hat
\sigma_{non-res}$ is free of mass singularities (terms proportional to 
$\ln(\hat s/m_f^2)$, where $m_f$ is the mass of an initial state or final
state fermion) and
on-shell singularities, i.e. logarithms of the form $\ln(|\hat s-M_W^2|)$.
It represents a gauge invariant subset of the complete ${\cal
O}(\alpha)$ contribution to the $W$ production process $q_i \bar
q_{i'} \to f\bar f'$, and its numerical impact can thus be studied
separately.  In Fig.~\ref{fig:nonres} we illustrate the effect of the
non-resonant contribution on the total parton level cross section by
showing the variation of the relative correction $\delta=\hat
\sigma_{non-res}/\hat \sigma^{(0)}$ (in percent) with $\sqrt{\hat s}$
where $\hat \sigma^{(0)}$ is the Born cross section,
\begin{equation}
\label{eq:lo}
\hat \sigma^{(0)} = {\pi\over 36}\,\alpha^2\left(1-{M_W^2\over
M_Z^2}\right)^{-2} 
{\hat s\over (\hat s-M_W^2)^2+({\hat s\over M_W}\,\Gamma_W)^2}~,
\end{equation}
in the $s$-dependent width scheme~\cite{Wackeroth:1996hz}. As can
be seen, the non-resonant contribution can be neglected in the
vicinity of the $W$ resonance but becomes increasingly important at
large center of mass energies due to the occurrence of large,
Sudakov-like logarithms of the form $(\alpha/ \pi) \ln^2(\hat
s/M_W^2)$. The kink at $\sqrt{\hat s}\approx 200$~GeV is due to the $WH$
threshold in the $W$ self-energy. 

The observable ${\cal O}(\alpha^3)$ cross section is obtained by
convoluting the parton cross section with the quark distribution
functions $f_{q/A} (x,Q^2)$ ($\hat s=x_1 x_2 s$) and summing over all
quark flavors $q$ and $q'$,
\begin{eqnarray}\label{eq:xsecobs}
d\sigma(s) &  =  & \sum_{q,\,q'} \int_0^1 d x_1 d x_2 \, \bigg(  f_{q/A}
(x_1,Q^2) f_{\bar{q'}/B} (x_2,Q^2)\, d\hat \sigma^{(0+1)} + (q
\leftrightarrow \bar{q'})  \bigg) 
\end{eqnarray}
with $(A,B)=(p,\bar{p})$ for the Tevatron and $(p,p)$ for the LHC.
The parton distribution functions 
depend on the QCD renormalization and factorization scales, $\mu_r$ and 
$\mu_f$, which we choose to be equal, $\mu_r=\mu_f=Q$.

\section{Phenomenological results}

\subsection{Preliminaries}
\label{sec:prelim}

We shall now discuss the phenomenological implications of the ${\cal
O}(\alpha)$ non-resonant EW corrections to $\ell\nu$ production at the
Tevatron ($p\bar p$ collisions at $\sqrt{s}=2$~TeV) and the LHC ($pp$ 
collisions at $\sqrt{s}=14$~TeV). For the numerical evaluation we chose the
following set of SM input parameters~\cite{caso}:
\begin{eqnarray}\label{eq:pars}
G_{\mu} = 1.16639\times 10^{-5} \; {\rm GeV}^{-2}, 
& \qquad & \alpha= 1/137.0359895, 
\nonumber \\ 
M_Z = 91.1867 \; {\rm GeV}, & \quad & 
\alpha_s\equiv\alpha_s(M_W^2)=0.121, 
\nonumber  \\
m_e  = 0.51099907 \; {\rm MeV}, &\quad &m_{\mu}=0.105658389 \; {\rm GeV},  
\quad m_{\tau}=1.777 \; {\rm GeV},
\nonumber  \\
m_u=0.0464 \; {\rm GeV}, & \quad & m_c=1.5 \; {\rm GeV}, 
\quad m_t=174 \; {\rm GeV},
\nonumber  \\
m_d=0.0465 \; {\rm GeV}, & \quad & m_s=0.15 \; {\rm GeV}, \quad m_b=4.7
\; {\rm GeV}, 
 \nonumber  \\
|V_{ud}|=|V_{cs}|=0.97 \; & \qquad & |V_{us}|=|V_{dc}|=0.22.
\end{eqnarray}
The fermion
masses only enter through loop contributions to the vector boson self
energies and as regulators of the collinear singularities which arise
in the calculation of the QED contribution. Non-zero light quark masses 
are only used in the calculation of the vector boson self energies. The 
light quark masses are
chosen such that the value for the hadronic contribution to the photon 
vacuum polarization for five active flavors, $\Delta
\alpha_{had}^{(5)}(M_Z^2)=0.028$~\cite{Eidelman:1995ny},
which is derived from low-energy $e^+ e^-$ data with the help of
dispersion relations, is recovered. $V_{ij}$ are the matrix elements of
the quark mixing matrix.

The $W$ mass and the Higgs boson mass, $M_H$, are related via loop 
corrections. A parametrization of the $W$ mass which, for 
$10~{\rm GeV}<M_H<1$~TeV, deviates by at most
0.5~MeV from the theoretical value including the full fermionic two-loop
contributions is given in Ref.~\cite{Awramik:2003rn}. Here we use the
somewhat older parametrization of Ref.~\cite{Degrassi:1998iy} 
\begin{eqnarray}
\label{eq:param}
M_W&=&M_W^0-0.0581 \; \ln\left(\frac{M_H}{100~{\rm GeV}}\right)
-0.0078 \ln^2\left(\frac{M_H}{100~{\rm GeV}}\right)
-0.085 \; \left(\frac{\alpha_s}{0.118}-1\right)
\nonumber \\[2.mm]
&-& 0.518 \; \left(\frac{\Delta \alpha_{had}^{(5)}(M_Z^2)}{0.028}-1\right)
+ 0.537 \; \left(\Bigl(\frac{m_t}{175~{\rm GeV}}\Bigr)^2-1\right)
\end{eqnarray}
with $M_W^0=80.3805$ GeV, which was used in the
analysis of the LEP data.  The parametrization of Eq.~(\ref{eq:param})
reproduces the result of Ref.~\cite{Degrassi:1998iy} to 0.2~MeV for 
$75~{\rm GeV}<M_H<350$~GeV. For the numerical discussion we choose
\begin{equation}
\label{eq:mhiggs}
M_H=120~{\rm GeV}, 
\end{equation}
which is consistent with current direct~\cite{Barate:2003sz} and
indirect bounds~\cite{lepewwg,clare}, and work in the $s$-dependent width
scheme. For the input parameters listed in Eq.~(\ref{eq:pars}) we obtain
$M_W=80.3612$~GeV. Higher order (irreducible) corrections connected with
the $\rho$-parameter are taken into account in our calculation by
replacing 
\begin{equation}
{\delta M_Z^2\over M_Z^2}\,-\,{\delta M_W^2\over M_W^2} \to
{\delta M_Z^2\over M_Z^2}\,-\,{\delta M_W^2\over M_W^2} -\Delta\rho^{HO}
\end{equation}
in the renormalized $Z$ and $\gamma,\, Z$ self-energies as described in
Appendix~B of Ref.~\cite{Baur:1999kt}. The $W$ width and the $q_i\bar
q_{i'}\to f\bar f'(\gamma)$ amplitude are calculated in the $G_\mu$
scheme. In the $G_\mu$ scheme, the fine structure constant, $\alpha$, is
replaced with 
\begin{equation}
\label{eq:replac}
\alpha \to {\sqrt{2}G_\mu M_W^2\over\pi}\left (1-{M_W^2\over
M_Z^2}\right ),
\end{equation}
and $M_W$ is obtained from Eq.~(\ref{eq:param}). For the $W$ width,
including ${\cal O}(\alpha)$ EW (see also Ref.~\cite{Denner:tx}) and QCD
corrections up to 
${\cal O}(\alpha_s^3)$~\cite{Baur:1999kt,Kataev,Kataev:1992dg}, we
obtain $\Gamma_W=2.0721$~GeV. 

In the numerical results presented below, we also take into account the
leading ${\cal O}(\alpha^2)$ weak corrections to resonant $W$ production
which can be obtained by
performing the following replacement in Eq.~(\ref{eq:res}):
\begin{equation} 
1+2 {\cal R}e(\tilde F_{weak}^{initial}+ \tilde 
F_{weak}^{final})(M_W^2) \to \left\vert \left(1+{1\over 2}\,
\tilde F_{weak}^{initial}(M_W^2)\right )\left(1+{1\over
2}\,\tilde F_{weak}^{final}(M_W^2) \right)\right\vert^2~.
\end{equation}

The dominant non-photonic electroweak corrections to resonant $W$
production can be taken into account in the effective Born
approximation (EBA) where the $W$ cross section, $\sigma^{EBA}$, is
calculated by replacing $\alpha$ in Eq.~(\ref{eq:lo}) with the
expression given in Eq.~(\ref{eq:replac}) and using the $W$ mass
obtained from Eq.~(\ref{eq:param}). In the following, we
shall use the EBA cross section as a reference when discussing how 
the ${\cal O}(\alpha)$ EW corrections affect physical observables.

Following a brief comparison with the calculation of
Ref.~\cite{Dittmaier:2001ay}, we discuss the impact of the non-resonant
EW corrections on the transverse mass distribution and how they affect
the value of the $W$-width extracted from that distribution. We then
consider the $W$ production 
cross section, the $W$ to $Z$ cross section ratio and the $W$ to $Z$
transverse mass ratio. To compute the hadronic cross section we use the
MRSR2 set of parton distribution 
functions~\cite{mrs}, and take the renormalization scale, $\mu_r$, and the
QED and QCD factorization scales, $\mu_{\rm QED}$ and $\mu_{\rm QCD}$, to be
$\mu_r^2=\mu_{\rm QED}^2=\mu_{\rm QCD}^2=M_W^2$. The detector acceptance
is simulated by imposing the following 
transverse momentum ($p_T$) and pseudo-rapidity ($\eta$) cuts:
\begin{equation}
p_T(\ell)>20~{\rm GeV,}\qquad\qquad |\eta(\ell)|<2.5, \qquad\qquad
\ell=e,\,\mu ,
\label{eq:lepcut}
\end{equation}
\begin{equation}
p\llap/_T>20~{\rm GeV,}
\label{eq:ptmisscut}
\end{equation}
where $p\llap/_T$ is the missing transverse momentum originating from
the neutrino. These cuts approximately model
the acceptance of the CDF II~\cite{cdfii} and D\O~\cite{dzero} detectors
at the Tevatron, and the ATLAS~\cite{atlas} and CMS~\cite{cms} detectors
at the LHC. Uncertainties in the energy measurements of the charged leptons 
in the detector are simulated in the calculation by Gaussian smearing 
of the particle four-momentum vector with standard deviation $\sigma$
which depends on the particle type and the detector. The numerical results 
presented here were calculated using $\sigma$ values based on the
D\O\ and ATLAS specifications. 

The granularity of the detectors and the size of the electromagnetic
showers in the calorimeter make it difficult to discriminate between 
electrons and photons with a small opening angle. In such cases we
recombine the four-momentum vectors of the electron 
and photon to an effective electron four-momentum vector. The exact
recombination procedure is detector dependent. For calculations
performed at Tevatron energies we use a procedure similar to
that used by the D\O\ Collaboration in Run~I, 
requiring that the electron and photon momentum four-vectors are
combined into an effective electron momentum four-vector if 
their separation in the pseudorapidity -- azimuthal angle
plane, 
\begin{equation}
\Delta R(e,\gamma)=\sqrt{(\Delta\eta(e,\gamma))^2+(\Delta\phi(e,
\gamma))^2}, 
\end{equation}
is $\Delta R(e,\gamma)<0.2$. For
$0.2<\Delta R(e,\gamma)<0.4$ events are rejected if
$E_{\gamma}>0.15 \; E_e$. Here $E_\gamma$ ($E_e$) is the energy of the
photon (electron) in the laboratory frame. For events with $0.2<\Delta 
R(e,\gamma)<0.3$ and 
$E_{\gamma}<0.15 \; E_e$, the electron and photon momentum four-vectors 
are again combined. At
LHC energies, we recombine the electron and photon four-momentum vectors
if $\Delta R(e,\gamma)<0.07$, similar to the resolution
expected for ATLAS~\cite{atlas}. Recombining the electron and photon
four-momentum vectors eliminates the mass singular logarithmic terms
originating from final state photon radiation and strongly reduces the
size of the QED-like final state corrections~\cite{Baur:1999kt}.

Muons are identified by hits in the muon chambers 
and the requirement that the associated track is consistent with a 
minimum ionizing particle. This limits the photon energy for small 
muon -- photon opening angles. For muons at the Tevatron, we again adopt
the D\O\ specifications and require that the energy 
of the photon is $E_{\gamma}<2$~GeV for $\Delta R(\mu,\gamma)<0.2$, and 
$E_{\gamma}<6$~GeV for $0.2<\Delta R(\mu,\gamma)<0.6$. At the LHC,
following Ref.~\cite{atlas}, we require  
the photon energy to be smaller than $E^\gamma_c=5$~GeV if 
$\Delta R(\mu,\gamma)<0.3$. The cut on the photon energy increases the
size of the photonic corrections for $m(\mu\nu)>100$~GeV~\cite{Baur:1999kt}.
For future 
reference, we summarize the lepton identification requirements in 
Table~\ref{tab:one}.
\begin{table}
\caption{Summary of lepton identification requirements. } 
\label{tab:one}
\vskip 5.mm
\begin{tabular}{cc}
\multicolumn{2}{c}{\bf Tevatron} \\
\multicolumn{1}{c}{electrons} & \multicolumn{1}{c}{muons} \\
\tableline
combine $e$ and $\gamma$ momentum four vectors if & reject events with 
$E_\gamma>2$~GeV \\ 
$\Delta R(e,\gamma)<0.2$ and if $E_\gamma<0.15~E_e$ for 
$0.2<\Delta R(e,\gamma)<0.3$ & for $\Delta R(\mu,\gamma)<0.2$ \\
\tableline
reject events with 
$E_\gamma>0.15~E_e$ & reject events with 
$E_\gamma>6$~GeV \\  
for $0.2<\Delta R(e,\gamma)<0.4$ & for $0.2<\Delta R(\mu,\gamma)<0.6$ \\
\tableline
\multicolumn{2}{c}{\bf LHC} \\
\multicolumn{1}{c}{electrons} & \multicolumn{1}{c}{muons} \\
\tableline
combine $e$ and $\gamma$ momentum four vectors if & reject events with 
$E_\gamma>5$~GeV \\ 
$\Delta R(e,\gamma)<0.07$ & for $\Delta R(\mu,\gamma)<0.3$ 
\end{tabular}
\end{table}

We use the input parameters listed in Eq.~(\ref{eq:pars}) and 
impose the cuts and lepton identification requirements described
above in all subsequent numerical simulations, unless explicitly 
noted otherwise. 

\subsection{Comparison with Ref.~[\ref{ditt}]}

As mentioned before, the matrix elements of the full ${\cal O}(\alpha)$
EW corrections to $q\bar q'\to\ell\nu+X$ were presented in
Ref.~\cite{Dittmaier:2001ay}, together with a discussion of how EW
radiative corrections 
influence the $p_T$ distribution of the charged lepton, and the
transverse mass distribution. In this section we compare the integrated
cross sections given in Ref.~\cite{Dittmaier:2001ay} for different ranges in
$p_T(\ell)$ at the Tevatron and LHC with the results of our calculation, using
the input parameters, PDFs, cuts, and the lepton -- photon 
recombination procedure of
Ref.~\cite{Dittmaier:2001ay}. The results are shown in
Table~\ref{tab:two}. 
\begin{table}
\caption{Integrated lowest order cross sections in the $G_\mu$ scheme,
$\sigma^{(0)}$, the relative
corrections for the pole approximation, $\delta_{PA}$, and
the full ${\cal O}(\alpha)$ EW corrections, $\delta$, for several ranges
of $p_T(\ell)$. Shown are the results of 
our calculation and of Ref.~[\ref{ditt}] for a) $p\bar p\to\ell\nu$ at
$\sqrt{s}=2$~TeV and b) $pp\to\ell\nu$ at $\sqrt{s}=14$~TeV. The statistical
uncertainties of the 
Monte Carlo integration are also shown. The lepton and photon momenta
are recombined for small values of $\Delta R(\ell,\gamma)$ using the
simplified procedure described in Ref.~[\ref{ditt}]. We also use the
same input parameters, PDFs and cuts as Ref.~[\ref{ditt}].} 
\label{tab:two}
\vskip 5.mm
\begin{tabular}{c||c|c|c|c|c|c}
\multicolumn{7}{c}{a) $p\bar p\to\ell\nu$, $\sqrt{s}=2$~TeV} \\
\tableline
$p_T(\ell)$ (GeV) & $25-\infty$ & $50-\infty$ & $75-\infty$ &
$100-\infty$ & $200-\infty$ & $300-\infty$ \\
\tableline
$\sigma^{(0)}$ (pb) Ref.~\cite{Dittmaier:2001ay} & 407.03(5) & 2.481(1) &
0.3991(1) & 0.1305(1) & 0.006020(2) & 0.0004821(1) \\
$\sigma^{(0)}$ (pb) this calc. & 407.02(7) & 2.4817(6) & 0.39926(9) &
0.13058(3) & 0.006017(2) &0.0004821(3) \\
\tableline
$\delta$ (\%)  Ref.~\cite{Dittmaier:2001ay} & $-1.8(1)$ & $-2.7(1)$ &
$-4.8(1)$ & $-6.3(1)$ & $-10.4(1)$ & $-13.6(1)$\\
$\delta$ (\%)  this calc. & $-1.7(1)$ & $-2.5(1)$ & $-4.7(1)$ & 
$-6.1(1)$ & $-10.1(1)$ & $-13.3(1)$ \\
\tableline
$\delta_{PA}$ (\%)  Ref.~\cite{Dittmaier:2001ay} & $-1.7(1)$ & $-1.6(1)$
& $-2.3(1)$ & $-2.5(1)$ & $-3.3(1)$ & $-3.9(1)$ \\
$\delta_{PA}$ (\%)  this calc. & $-1.7(1)$ & $-1.5(1)$ & $-2.2(1)$ & 
$-2.4(1)$ & $-3.1(1)$ & $-3.7(1)$\\
\tableline
\tableline
\multicolumn{7}{c}{b) $pp\to\ell\nu$, $\sqrt{s}=14$~TeV} \\
\tableline
$p_T(\ell)$ (GeV) & $25-\infty$ & $50-\infty$ & $100-\infty$ &
$200-\infty$ & $500-\infty$ & $1000-\infty$ \\
\tableline
$\sigma^{(0)}$ (pb) Ref.~\cite{Dittmaier:2001ay} & 1933.5(3) & 11.50(1) &
0.8198(4) & 0.1015(1) & 0.005277(1) & 0.0003019(1) \\
$\sigma^{(0)}$ (pb) this calc. & 1933.4(3) & 11.499(2) & 0.8202(1) &
0.10155(2) & 0.005277(1) &0.0003019(1) \\
\tableline
$\delta$ (\%)  Ref.~\cite{Dittmaier:2001ay} & $-1.8(1)$ & $-2.7(1)$ &
$-6.2(1)$ & $-10.2(1)$ & $-19.6(1)$ & $-29.6(1)$\\
$\delta$ (\%)  this calc. & $-1.8(1)$ & $-2.3(1)$ & $-6.0(1)$ & 
$-10.1(1)$ & $-19.1(1)$ & $-28.6(1)$ \\
\tableline
$\delta_{PA}$ (\%)  Ref.~\cite{Dittmaier:2001ay} & $-1.8(1)$ & $-1.5(1)$
& $-1.6(1)$ & $-1.6(1)$ & $-2.4(1)$ & $-3.6(1)$ \\
$\delta_{PA}$ (\%)  this calc. & $-1.8(1)$ & $-1.2(1)$ & $-1.5(1)$ & 
$-1.6(1)$ & $-2.4(1)$ & $-3.4(1)$
\end{tabular}
\end{table}
The table lists the lowest order cross section in the $G_\mu$ scheme,
$\sigma^{(0)}$, and the relative corrections,
\begin{equation}
\delta={\sigma^{{\cal O}(\alpha^3)}-\sigma^{(0)}\over\sigma^{(0)}}
\end{equation}
for the full ${\cal O}(\alpha)$ EW corrections, and
\begin{equation}
\delta_{PA}={\sigma^{PA}-\sigma^{(0)}\over\sigma^{(0)}}
\end{equation}
for the ${\cal O}(\alpha)$ EW corrections in the pole approximation. Our
results are found to agree within the statistical accuracy of the Monte
Carlo integration with those obtained in Ref.~\cite{Dittmaier:2001ay}
over the entire lepton $p_T$ range. 

\subsection{Non-resonant corrections to the transverse mass
distribution}

Table~\ref{tab:two} shows that the non-resonant EW corrections quickly
become important for transverse momenta above the Jacobian peak region,
$p_T(\ell)>M_W/2\approx 40$~GeV. The deviation of the ${\cal
O}(\alpha^3)$ cross section in the pole approximation from the full NLO
result at large $p_T(\ell)$ is due to large Sudakov-like
electroweak logarithms of the form $(\alpha/\pi)\ln^2(m(\ell\nu)/M_V)$
($V=W,\, 
Z$)~\cite{cc} which mostly arise from the contribution of the $W,Z$ box
diagrams and to a lesser extend from the energy dependence of the form
factors $F_{weak}^a$ and $F_\gamma-\sum_a\tilde F^a_{YFS}$. The
same qualitative behavior is also expected for the transverse mass
distribution, which is used to determine the $W$ mass and
width~\cite{unknown:2003sv}. The transverse mass is
defined by
\begin{equation}
M_T=\sqrt{2p_T(\ell)p_T(\nu)(1-\cos\phi^{\ell\nu})}~,
\label{eq:mt}
\end{equation}
where $p_T(\nu)$ is the transverse momentum of the neutrino, and
$\phi^{\ell\nu}$ is the angle between the 
charged lepton and the neutrino in the transverse plane. The neutrino 
transverse momentum is identified with the missing transverse momentum,
$p\llap/_T$, in the event. 

\begin{figure}[t!]
\begin{center}
\includegraphics[width=16cm]{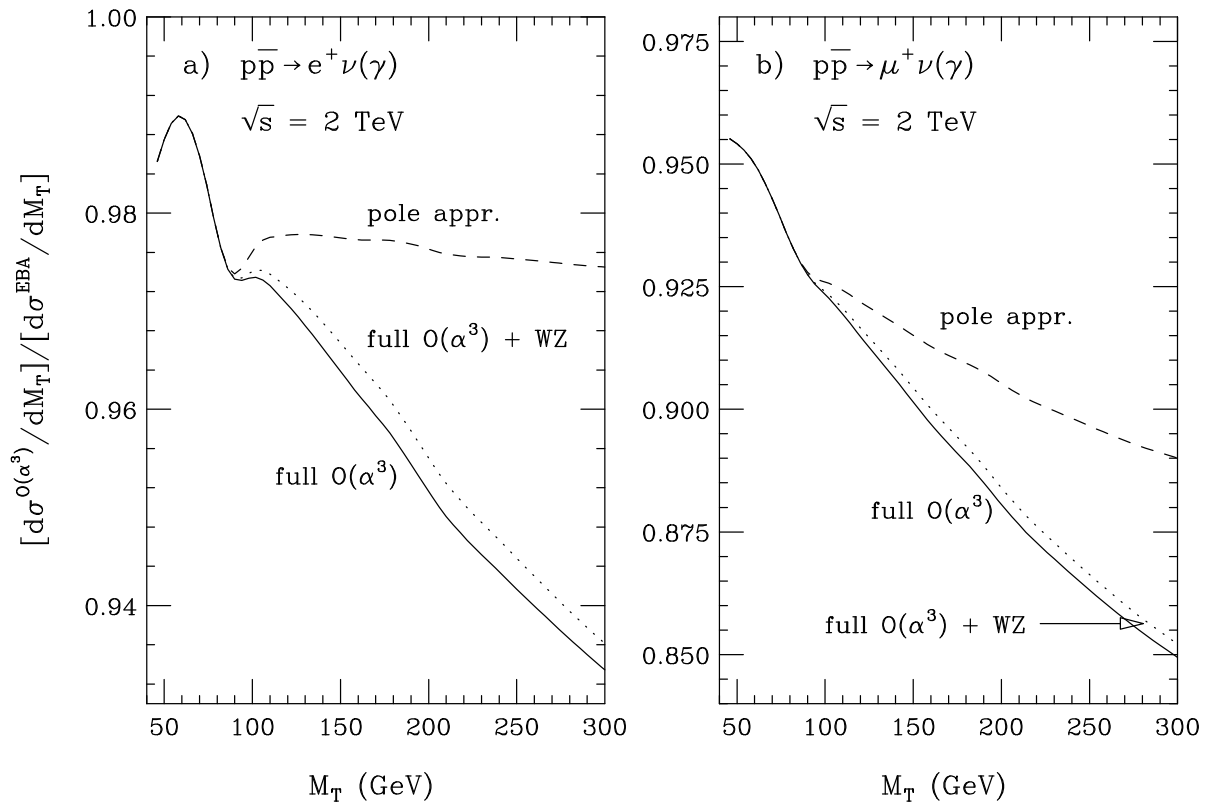}
\vspace*{2mm}
\caption{\label{fig:one}
The ratio $[d\sigma^{{\cal O}(\alpha^3)}/dM_T]/[d\sigma^{
EBA}/dM_T]$ as a function of the transverse mass for a) $p\bar p\to
e^+\nu_e(\gamma)$ and b) $p\bar p\to\mu^+\nu_\mu(\gamma)$ at
$\sqrt{s}=2$~TeV. The solid lines show the ratio of
the complete ${\cal O}(\alpha^3)$ electroweak and the EBA differential
cross section. The dashed lines display the corresponding ratio for the 
case where only the resonant ${\cal O}(\alpha)$ EW corrections (see
Eq.~(\ref{eq:res})) are taken into account (pole approximation). The
dotted lines show the ratio 
when the $p\bar p\to W^+(\to\ell\nu)Z(\to\bar\nu\nu)$ background is
included in addition to the complete ${\cal O}(\alpha)$ EW
corrections. The
cuts and lepton identification requirements imposed are described in
Sec.~\ref{sec:prelim}. 
}\vspace{-7mm}
\end{center}
\end{figure}

The ratio of the complete ${\cal O}(\alpha^3)$ electroweak and the EBA
differential cross section as a function of $M_T$ is shown in
Fig.~\ref{fig:one}. 
In order to 
make the effect of the non-resonant weak corrections more
transparent, we also show the corresponding ratio for the case of
the ${\cal O}(\alpha)$ EW corrections in the pole approximation (dashed
lines)~\cite{Baur:1999kt}. For $M_T\leq M_W$, the pole approximation is
seen to very well 
represent the complete electroweak ${\cal O}(\alpha)$ corrections. In this
region, the shape change in the $M_T$ distribution is largely due to the
contribution of the final state QED-like corrections. Due to the
recombination of electrons and photons, the EW corrections in the pole
approximation reduce the $e\nu(\gamma)$ differential cross section by
only $2-3\%$ 
over the transverse mass region considered. In the muon case, the cut on
the photon energy for photons which have a small opening angle with the
muon reduces the hard photon part of the ${\cal O}(\alpha^3)$
$\mu\nu(\gamma)$ cross section. As a result, the QED-like corrections are
much more pronounced and display a much stronger dependence on the
transverse mass than in the electron case. Without taking the lepton
identification criteria of Table~\ref{tab:one} into account, the
QED-like corrections in the electron case are larger, due to the mass
singular logarithmic corrections which originate from final state photon
radiation. The non-resonant EW
corrections are seen to increase rapidly in size with $M_T$. For
$M_T=300$~GeV, they reduce the cross section by about 4\% at
the Tevatron. The dotted lines, finally, show cross section ratio
when the $p\bar p\to W^+(\to\ell^+\nu)Z(\to\bar\nu\nu)$ background is
included in addition to the complete ${\cal O}(\alpha)$ EW
corrections. The $WZ$ background is seen to be much smaller than the 
${\cal O}(\alpha)$ EW corrections.

\begin{figure}[t!]
\begin{center}
\includegraphics[width=16cm]{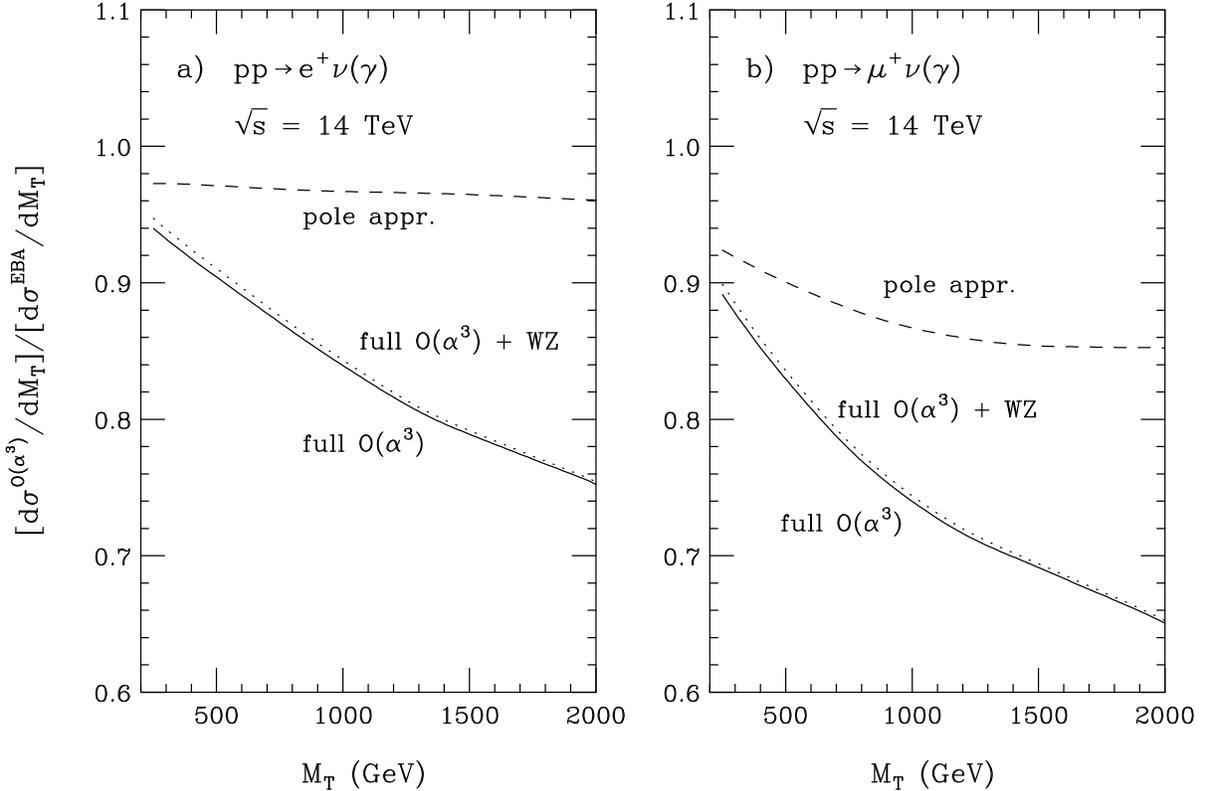}
\vspace*{2mm}
\caption{\label{fig:two}
The ratio $[d\sigma^{{\cal O}(\alpha^3)}/dM_T]/[d\sigma^{
EBA}/dM_T]$ as a function of the transverse mass for a) $pp\to
e^+\nu_e(\gamma)$ and b) $pp\to\mu^+\nu_\mu(\gamma)$ at
$\sqrt{s}=14$~TeV. The solid lines show the ratio of
the complete ${\cal O}(\alpha^3)$ electroweak and the EBA differential
cross section. The dashed lines display the corresponding ratio for the 
case where only the resonant ${\cal O}(\alpha)$ EW corrections (see
Eq.~(\ref{eq:res})) are taken into account (pole approximation). The
dotted lines show the ratio 
when the $pp\to W^+(\to\ell^+\nu)Z(\to\bar\nu\nu)$ background is
included in addition to the complete ${\cal O}(\alpha^3)$ EW
corrections. The
cuts and lepton identification requirements imposed are described in
Sec.~\ref{sec:prelim}. 
}\vspace{-7mm}
\end{center}
\end{figure}

Figure~\ref{fig:two} shows the corresponding results for the
LHC in the high transverse mass tail.
For $M_T>1.5$~TeV, the non-resonant ${\cal O}(\alpha)$ EW corrections
reduce the differential cross section by 20\% or more and thus are of the
same size as the ${\cal O}(\alpha_s)$ corrections. This is 
larger than the expected statistical uncertainty in a 200~GeV bin centered
at $M_T=1.5$~TeV for 100~fb$^{-1}$. It will thus be important to
take into account the non-resonant weak corrections when searching for
new heavy $W$ bosons, such as Kaluza-Klein excitations of the $W$
appearing in TeV-scale models with extra dimensions~\cite{kk}, at the
LHC. The results
shown in Fig.~\ref{fig:two}, however, should be interpreted with 
caution. Since the non-resonant weak corrections become large for 
transverse masses above 1~TeV, they need to be resummed in order to
obtain accurate predictions in this phase space region (for a
recent review of the resummation of electroweak Sudakov-like
logarithms see Ref.~\cite{melles}). Although the resummation of
electroweak Sudakov-like logarithms in general four fermion electroweak
processes has been discussed in the literature~\cite{penin}, 
a calculation of $\ell\nu$ 
production in hadronic collisions which includes resummation of electroweak
logarithms has not been carried out yet. 

\subsection{Non-resonant EW radiative corrections and the $W$ width}

\begin{figure}[t!]
\begin{center}
\includegraphics[width=16cm]{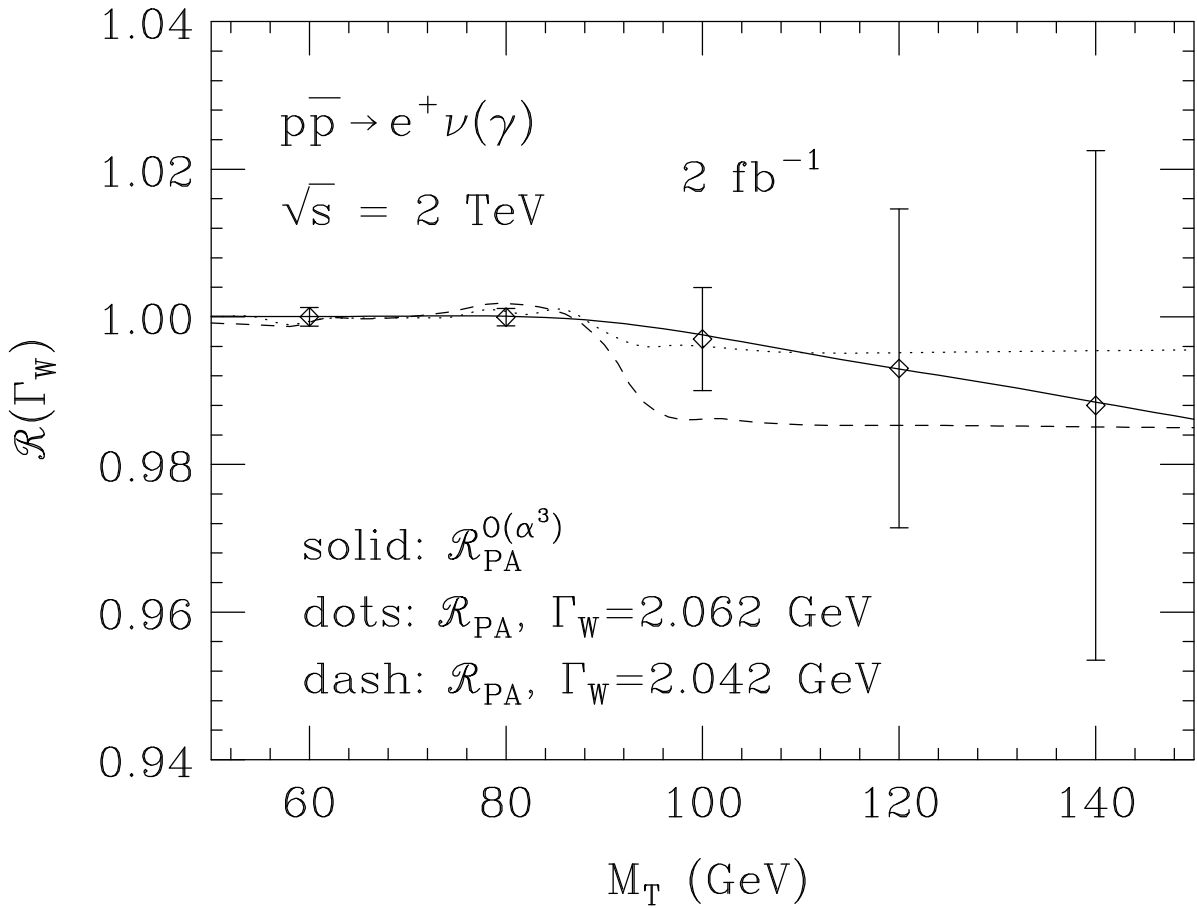}
\vspace*{2mm}
\caption{\label{fig:three}
The ratio ${\cal R}(\Gamma_W)$ (see text for definition) as a
function of $M_T$ for $p\bar p\to e^+\nu(\gamma)$ at the
Tevatron. Results are shown for 
$\Gamma_W=\Gamma_W^{SM}-10~{\rm MeV}=2.062$~GeV (dotted line) and 
$\Gamma_W=\Gamma_W^{SM}-30~{\rm MeV}=2.042$~GeV (dashed line) with the
cross sections calculated in the pole approximation. The solid
line displays the ratio ${\cal R}^{{\cal
O}(\alpha^3)}_{PA}=[d\tilde\sigma^{{\cal O}(\alpha^3)}/dM_T]/
[d\tilde\sigma^{PA}/dM_T]$ in the SM. The data points and 
error bars indicate the measurements and statistical
uncertainties expected for ${\cal R}^{{\cal O}(\alpha^3)}_{PA}$ for
20~GeV bins and an
integrated luminosity of 2~fb$^{-1}$, assuming that the data are
described by the SM prediction 
including the full ${\cal O}(\alpha)$ EW corrections. The
cuts and lepton identification requirements imposed are described in
Sec.~\ref{sec:prelim}. 
}\vspace{-7mm}
\end{center}
\end{figure}

We now discuss how the non-resonant ${\cal O}(\alpha)$ EW corrections
affect the $W$ width determined from the high transverse mass tail. In
Run~I of the Tevatron, $\Gamma_W$ has been measured by the CDF and D\O\
Collaborations  using this technique with a combined 
uncertainty of 105~MeV~\cite{unknown:2003sv}. In 
Run~II, with an integrated luminosity of
2~fb$^{-1}$, one expects to achieve a precision of 50~MeV per lepton
channel and experiment~\cite{Brock:1999ep}. Assuming that the error 
correlation of CDF and D\O\ data, as in Run~I, is 
small~\cite{unknown:2003sv} for the $W$-width measurement, this results
in an expected overall   
uncertainty of $\delta\Gamma_W=25-30$~MeV. In the experimental
analysis, the measured transverse mass distribution is compared with the
theoretical prediction for various values of the $W$ width where the
total cross section has been normalized to the experimental
value~\cite{Affolder:2000mt,Abazov:2002xj}. This is equivalent to
analyzing the normalized $M_T$ distribution,
\begin{equation}
{d\tilde\sigma\over dM_T}(\Gamma_W)={1\over\sigma_{tot}(\Gamma_W)}
~{d\sigma\over dM_T}(\Gamma_W), 
\end{equation}
where $\sigma_{tot}$ is the total $\ell\nu(\gamma)$ cross section within
cuts. At lowest 
order, the $W$ width enters the cross section in the form of the squared
$W$ propagator (see Eq.~(\ref{eq:lo})),
\begin{equation}
|D_W(\hat s)|^2=[(\hat s-M_W^2)^2+\hat s^2\Gamma_W^2/M_W^2]^{-1}.
\end{equation}
The lowest order total $\ell\nu$ production cross section thus is
proportional to $1/\Gamma_W$. As a result, $d\tilde\sigma/dM_T$ is
proportional to $\Gamma_W$ for $M_T\gg M_W$. This is clearly displayed in
the ratio of the normalized $M_T$ distribution for arbitrary $\Gamma_W$
to the normalized $M_T$ distribution in the SM,
\begin{equation}
{\cal R}(\Gamma_W)=\frac{\displaystyle\frac{d\tilde\sigma}{dM_T}(\Gamma_W)}{
\displaystyle\frac{d\tilde\sigma}{dM_T}(\Gamma^{SM}_W)} ~,
\end{equation}
where $\Gamma_W^{SM}=2.072$~GeV is the SM $W$ width (see
Sec.~\ref{sec:prelim}). ${\cal R}(\Gamma_W)$ in the pole approximation,
${\cal R}_{PA}(\Gamma_W)$, is shown in Fig.~\ref{fig:three} for
$p\bar p\to e^+\nu(\gamma)$ at the Tevatron with 
$\Gamma_W=\Gamma_W^{SM}-10~{\rm MeV}=2.062$~GeV (dotted line) and 
$\Gamma_W=\Gamma_W^{SM}-30~{\rm MeV}=2.042$~GeV (dashed line). 
Qualitatively similar results are
obtained for the muon final state and at LHC energies. ${\cal
R}_{PA}(\Gamma_W)$ is seen to be almost constant and approximately equal to
$\Gamma_W/\Gamma_W^{SM}$ for $M_T>100$~GeV. For $M_T<M_W$, the ratio is
almost independent of $\Gamma_W$, and very close to one. If no detector
resolution effects are taken into account, $d\tilde\sigma/dM_T$ is
proportional to $1/\Gamma_W$ for $M_T=M_W$ at lowest order, resulting in
an enhancement proportional to $\Gamma_W^{SM}/\Gamma_W$ in ${\cal
R}(\Gamma_W)$. However, detector resolution effects largely dilute
this effect. Above the $W$ mass, ${\cal R}_{PA}(\Gamma_W)$ rapidly drops
from ${\cal R}_{PA}(\Gamma_W)\approx 1$ to ${\cal
R}_{PA}(\Gamma_W)\approx\Gamma_W/\Gamma_W^{SM}$. The $M_T$ range over
which the transition occurs sensitively depends on the missing
transverse momentum resolution of the detector.

The solid line in Fig.~\ref{fig:three} shows the ratio 
\begin{equation}
{\cal R}^{{\cal
O}(\alpha^3)}_{PA}=\frac{\displaystyle\frac{d\tilde\sigma^{{\cal
O}(\alpha^3)}}{dM_T}}{
\displaystyle\frac{d\tilde\sigma^{PA}}{dM_T}} ~,
\end{equation}
for $\Gamma_W=\Gamma_W^{SM}$, 
where $d\tilde\sigma^{{\cal O}(\alpha^3)}/dM_T$ and 
$d\tilde\sigma^{PA}/dM_T$ are the SM normalized $M_T$ 
distributions for the complete ${\cal O}(\alpha^3)$ calculation and in
the pole approximation, respectively. The data points and 
error bars in Fig.~\ref{fig:three} indicate the measurements and statistical
uncertainties expected for ${\cal R}^{{\cal O}(\alpha^3)}_{PA}$ for
20~GeV bins and an
integrated luminosity of 2~fb$^{-1}$, assuming that the data are
described by the SM prediction 
including the full ${\cal O}(\alpha)$ EW corrections. Due to the
non-resonant EW 
corrections, ${\cal R}^{{\cal O}(\alpha^3)}_{PA}$ gradually decreases
for $M_T>90$~GeV, the region used by the Tevatron experiments to extract
the $W$ width from Run~I data~\cite{Affolder:2000mt,Abazov:2002xj}.
Although the shape change of the transverse mass 
distribution due to the non-resonant ${\cal O}(\alpha)$ EW correction
differs significantly from that caused by a non-standard $W$ width, the
expected statistical uncertainties make it difficult to distinguish
between a small negative shift in $\Gamma_W$ and the effect of
non-resonant ${\cal O}(\alpha)$ EW corrections. A $\chi^2$ analysis
shows that, were the non-resonant ${\cal O}(\alpha)$ EW corrections
ignored, the value of the $W$ width extracted from the $M_T>90$~GeV data
region would be shifted by
\begin{equation}
\Delta\Gamma_W=-7.2~{\rm MeV.}
\end{equation}
Since the $M_T$ distribution depends little on the detector resolution
for $M_T>90$~GeV, the shift in $\Gamma_W$ is almost independent of these
effects~\cite{Brock:1999ep}. For the precision of $\Gamma_W$ expected
in Run~II, a difference of $\approx 7$~MeV in the extracted value of the
$W$ width cannot be ignored, and the complete ${\cal O}(\alpha^3)$
calculation should be used to compare theory and data.

\subsection{Non-resonant corrections to the $W$ boson cross section and
the $W$ to $Z$ cross section ratio}

As mentioned in the Introduction, the $W$ width can also be determined
from the cross section ratio $R_{W/Z}$ (see Eq.~(\ref{eq:ratio})),
together with the theoretical prediction for the ratio of the total $W$
and $Z$ production cross sections, the LEP measurement of the $Z\to
\ell^+\ell^-$ branching ratio, and the SM prediction for the
$W\to\ell\nu$ decay width. Since the QCD corrections to $W$ and $Z$ production
are very similar, they cancel almost perfectly in the $W$ to $Z$ cross
section ratio; the ${\cal O}(\alpha_s)$ corrections to $R_{W/Z}$ are of
${\cal O}(1\%)$ or less, depending on the set of parton distribution
functions used~\cite{willy}. In addition many experimental uncertainties,
such as the luminosity uncertainty, cancel in the cross section ratio. 
%The $W$ boson cross section can also be used to determine the integrated
%luminosity at the Tevatron and LHC. 
Accurate knowledge of how electroweak corrections affect $R_{W/Z}$ is thus
very important. 

The $W$ cross section may be used as a luminosity monitor in the
future~\cite{CDFZSIGN,DITT1}. This requires that the $W$ cross section
is reliably computed with small uncertainty. In order to achieve this,
it is essential to know how EW radiative corrections affect the
$W\to\ell\nu$ cross section.

The size of the ${\cal O}(\alpha)$ electroweak
corrections to the total $p\bar p\to\ell\nu X$ cross section and to 
$R_{W/Z}$ is 
sensitive to the acceptance cuts and whether lepton identification
requirements are taken into account or not. In Table~\ref{tab:three}, we 
list the electroweak $K$-factor,
\begin{equation}
K^{EW}={\sigma^{{\cal O}(\alpha^3)}(p\bar p\to
W\to\ell\nu X)\over\sigma^{EBA}(p\bar p\to W\to\ell\nu)}~, 
\end{equation}
and the correction factor for $R_{W/Z}$,
\begin{equation}
K_R^{EW}={R_{W/Z}^{{\cal O}(\alpha^3)}\over R_{W/Z}^{EBA}}~,
\end{equation}
for the acceptance cuts listed in Eqs.~(\ref{eq:lepcut})
and~(\ref{eq:ptmisscut}) with and without taking the lepton identification
requirements of Table~\ref{tab:one} into account. 
\begin{table}
\caption{The electroweak $K$-factor $K^{EW}=\sigma^{{\cal
O}(\alpha^3)}(p\bar p\to W\to\ell\nu X)/\sigma^{EBA}(p\bar p\to 
W\to\ell\nu)$ ($\ell=e,\,\mu$) and the correction factor to $R_{W/Z}$, 
$K_R^{EW}=R_{W/Z}^{{\cal O}(\alpha^3)}/R_{W/Z}^{EBA}$, with 
\protect{$75~{\rm GeV}<m(\ell^+\ell^-)\break <105$}~GeV, for $p\bar p$
collisions at $\protect{\sqrt{s}=2}$~TeV. Shown are the predictions 
without and with the lepton identification requirements of
Table~\protect{\ref{tab:one}} taken into account. The cuts imposed are 
listed in Eqs.~(\protect{\ref{eq:lepcut}})
and~(\protect{\ref{eq:ptmisscut}}). The energy and momentum resolutions
used are described in Sec.~\protect{\ref{sec:prelim}}. }
\label{tab:three}
\vskip 5.mm
\begin{tabular}{ccc}
\multicolumn{1}{c}{} &
\multicolumn{1}{c}{without lepton id.} &
\multicolumn{1}{c}{with lepton id.} \\
\multicolumn{1}{c}{} &
\multicolumn{1}{c}{requirements} &
\multicolumn{1}{c}{requirements} \\
\tableline
$K^{EW}~(p\bar p\to e^+\nu X)$, full ${\cal O}(\alpha^3)$ & 0.963 & 0.984 \\
$K^{EW}~(p\bar p\to e^+\nu X)$, pole appr. & 0.963 & 0.984 \\
$K^{EW}~(p\bar p\to\mu^+\nu X)$, full ${\cal O}(\alpha^3)$ & 0.979 & 0.944 \\
$K^{EW}~(p\bar p\to\mu^+\nu X)$, pole appr. & 0.978 & 0.943 \\
\tableline
$K^{EW}_R~(e)$, full ${\cal O}(\alpha^3)$ & 1.024 & 0.992 \\
$K^{EW}_R~(\mu)$, full ${\cal O}(\alpha^3)$ & 1.002 & 1.045 \\
\end{tabular}
\end{table}
For $K^{EW}$, we also
list the result obtained in the pole approximation. To compute
the ${\cal O}(\alpha^3)$ $Z$ boson cross section entering $R_{W/Z}$, we
use the full ${\cal O}(\alpha^3)$ calculation of di-lepton production in
hadronic collisions described in Ref.~\cite{Baur:2001ze}. We 
include photon exchange and 
$\gamma Z$ interference effects, and impose a cut on the di-lepton 
invariant mass of $75~{\rm GeV}<m(\ell^+\ell^-)<105~{\rm GeV}$. The
definition of the effective Born approximation in the $Z$ case is given
in Ref.~\cite{Baur:2001ze}. The
values listed in Table~\ref{tab:three} update the results presented in
Ref.~\cite{Baur:1999kt} which were obtained using the pole approximation
for the $W\to\ell\nu$ cross section and only included QED corrections in
the $Z$ boson case.

From the results listed in Table~\ref{tab:three} one observes that the
non-resonant ${\cal O}(\alpha)$ EW corrections have a very small effect
on the $W$ boson cross section; they change the electroweak $K$ factor
by ${\cal O}(10^{-3})$ or less. The full ${\cal O}(\alpha)$ EW
corrections decrease the $W$ cross section and increase $R_{W/Z}$ by
several per cent for the cuts imposed. When lepton
identification requirements are included, the 
corrections are reduced in the electron case and enhanced in the muon
case. Unlike the QCD corrections, the electroweak corrections do not
cancel in $R_{W/Z}$. In $Z\to\ell^+\ell^-$ both leptons can emit
photons, whereas only the 
charged lepton radiates in $W\to\ell\nu$ decays. Since final state
photonic corrections are the dominating contribution to the ${\cal
O}(\alpha)$ EW corrections, the ${\cal O}(\alpha)$ corrections to the
$W$ and $Z$ cross sections are quite different, and therefore do not 
cancel in $R_{W/Z}$. They are of the same size as the QCD corrections to
$R_{W/Z}$. 
The size of the ${\cal O}(\alpha)$ EW corrections to the total $p\bar
p\to\ell\nu X$ cross section and to $R_{W/Z}$ is similar to the
statistical uncertainty from 72~pb$^{-1}$ of data from
Run~II~\cite{run2r}. It is thus important to take the EW
radiative corrections into account in the Run~II data analysis.

\subsection{The full ${\cal O}(\alpha)$ electroweak corrections to the
$W$ to $Z$ transverse mass ratio}

Since detectors cannot directly 
detect the neutrinos produced in the leptonic $W$ boson decays,
$W\to\ell\nu$, and cannot measure the
longitudinal component of the recoil momentum, there is insufficient
information to reconstruct the invariant mass of the $W$ boson. 
Instead, the transverse mass distribution of the final state
lepton pair, or the transverse momentum distribution of the charged
lepton are used~\cite{cdfwmass,d0wmass} to extract $M_W$. The $M_T$
distribution has the advantage of being invariant under transverse
boosts to first order in the velocity of the $W$ boson. On the other
hand, the transverse mass depends on an accurate reconstruction of the
neutrino direction which leads to a set of experimental requirements which
are difficult in practice to control~\cite{Brock:1999ep}. 

The $W$ mass can also be determined from the ratio of the transverse
mass distributions of the $W$ and $Z$ boson~\cite{MR,gike,shpakov}. 
The advantage of this method is that one can cancel common
scale factors in ratios and directly determine ${M_{W}/M_{Z}}$, 
which can be compared with the precise value of $M_Z$ from the LEP
experiments. The downside of the ratio method is that the statistical
precision of the $Z$ sample is directly propagated 
into the resultant overall uncertainty of $M_{W}$. At high luminosities,
or when 
a detailed understanding of the detector response is not available, the
transverse mass ratio of $W$ to $Z$ bosons offers advantages in
determining the $W$ mass over the $M_T$ distribution~\cite{Brock:1999ep}.

The
transverse mass ratio of $W$ and $Z$ bosons is defined as
\begin{equation}
R_{M_T}(X_{M_T})={A_W(X^W_{M_T}=X_{M_T})\over A_Z(X^Z_{M_T}=X_{M_T})}~,
\end{equation}
where $A_V$ ($V=W,\,Z$) is the differential cross section
\begin{equation}
A_V(X^V_{M_T})={d\sigma_V\over dX^V_{M_T}}
\end{equation}
with respect to the scaled transverse mass,
\begin{equation}
X^V_{M_T}={M^V_T\over M_V}~.
\end{equation}
The transverse mass of the lepton pair in $Z$ boson events is defined in
complete analogy to Eq.~(\ref{eq:mt}):
\begin{equation}
M^Z_T=\sqrt{2p_T(\ell^+)p_T(\ell^-)(1-\cos\phi^{\ell\ell})}~,
\end{equation}
where $\phi^{\ell\ell}$ is the angle between the two charged leptons in the 
transverse plane. The ${\cal O}(\alpha)$ EW radiative corrections to
$R_{M_T}$ were calculated in Ref.~\cite{Baur:1999kt} in the
approximation where only the QED corrections were taken into account for
$Z\to\ell^+\ell^-$ and the pole approximation was used for
$W\to\ell\nu$. Here we present results which include the complete ${\cal
O}(\alpha)$ EW radiative corrections. 

\begin{figure}[t!]
\begin{center}
\includegraphics[width=16cm]{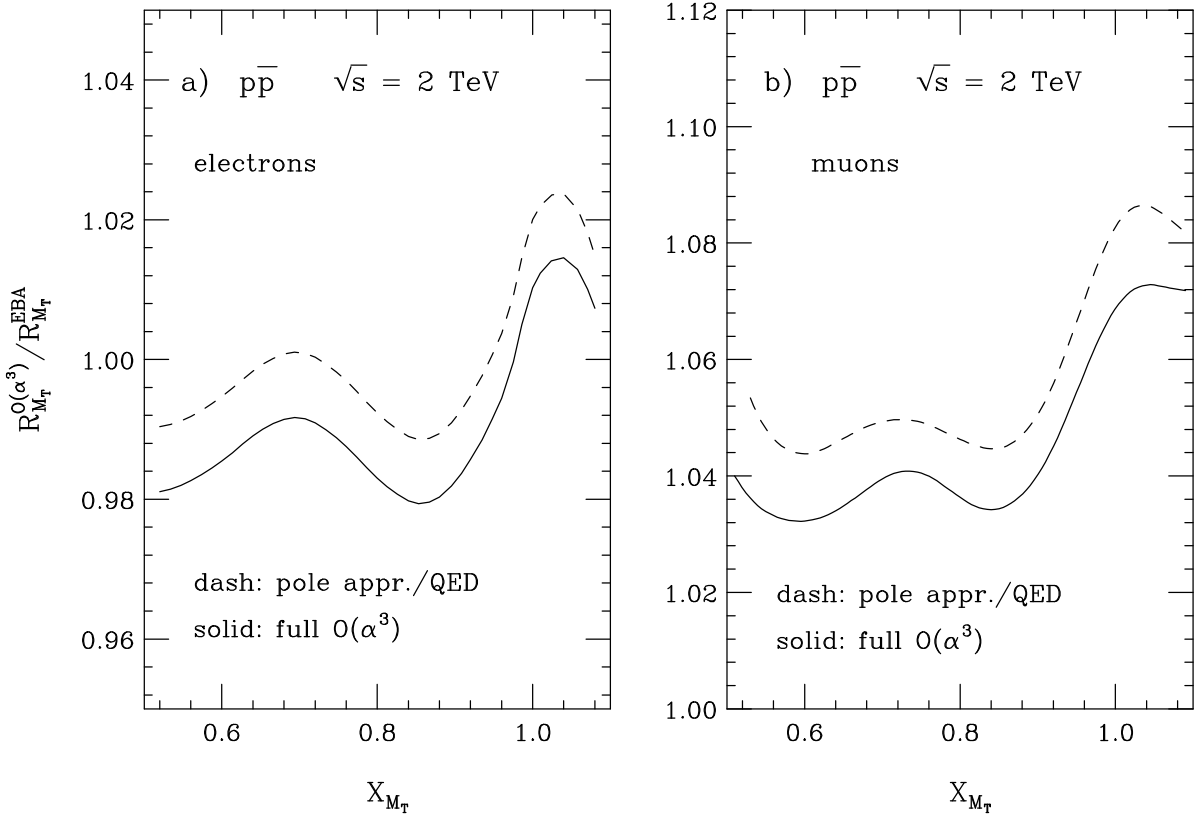}
\vspace*{2mm}
\caption{\label{fig:four}
Ratio of the $\protect{{\cal O}(\alpha^3)}$ and the $W^+$ to
$Z$ transverse mass ratio in the effective Born approximation as a
function of the scaled transverse mass, 
$X_{M_T}$, for $\protect{p\bar p}$ collisions at $\protect{\sqrt{s}=2}$~TeV, a)
for electron, and b) for muon final states. The solid 
lines show the result when the complete ${\cal O}(\alpha)$ EW radiative
corrections are taken into account. The dashed lines give the result for
the EW radiative corrections to $W$ production in the pole
approximation, and when only the QED corrections to $Z$ boson production
are taken into account (as in Ref.~[\ref{99pap}]). 
The cuts and lepton identification requirements imposed are described in
Sec.~\protect{\ref{sec:prelim}}. For
$\protect{p\bar p\to\ell^+\ell^-(\gamma)}$, we in addition require the
di-lepton  
invariant mass to satisfy the constraint $\protect{75~{\rm
GeV}<m(\ell^+\ell^-)<105~{\rm GeV}}$. }
\vspace{-7mm}
\end{center}
\end{figure}

The ratio of the ${\cal O}(\alpha^3)$ and the $W$ to $Z$ transverse
mass ratio in the effective Born
approximation is shown in Fig.~\ref{fig:four}. To calculate the ${\cal
O}(\alpha)$ electroweak corrections to $Z$ boson production, we again
use the results of Ref.~\cite{Baur:2001ze}. 
As before, photon exchange and $\gamma Z$ interference effects are
included and an additional cut on the di-lepton 
invariant mass of $75~{\rm GeV}<m(\ell^+\ell^-)<105~{\rm GeV}$ has been
imposed. The complete ${\cal O}(\alpha)$ EW radiative corrections (solid
lines) are uniformly about 1\% smaller than those obtained when the pole
approximation is used for $W$ production and only QED corrections to $Z$
boson production are taken into account (dashed lines). Most of this
effect originates from the genuine weak corrections to the $Z$ boson
cross section. In the electron case, the 
${\cal O}(\alpha)$ EW corrections change $R_{M_T}$ by $1-2\%$. They are
larger in the muon case ($3-7\%$) due to the lepton identification
requirements which reduce the hard photon part of the ${\cal
O}(\alpha^3)$ cross section and thus enhance the effect of the virtual
corrections. In the resonance region, $X_{M_T}\approx 1$, hard photonic
corrections reduce both the $W$ and $Z$ boson cross sections. As mentioned
before, in $p\bar p\to\ell\nu(\gamma)$ only one of the two leptons can  
emit a photon, whereas both leptons in $p\bar p\to\ell^+\ell^-(\gamma)$ 
can radiate. As a result, the effect of the hard photonic corrections is
more pronounced in the $Z$ case, resulting in a resonance-like
enhancement of $R_{M_T}$ for $X_{M_T}\approx 1$.

\section{Conclusions}

We have presented a calculation of the ${\cal O}(\alpha)$ corrections to
$p\,p\hskip-7pt\hbox{$^{^{(\!-\!)}}$} \to W^\pm\to\ell^\pm\nu$ based on the
complete set of one-loop Feynman diagrams contributing to $\ell\nu$
production, using the methods developed in
Ref.~\cite{Wackeroth:1996hz}. The calculation is based on a
combination of analytic and Monte Carlo integration techniques. Lepton
mass effects are included in the approximation where only mass singular 
terms originating from the collinear singularity associated with final
state photon radiation are retained. The ultraviolet
divergences associated with the virtual corrections are regularized
using dimensional regularization and the {\sc on-shell} renormalization 
scheme~\cite{bo86}. The cross sections obtained using our matrix
elements were found to be in
very good agreement with those obtained in Ref.~\cite{Dittmaier:2001ay}.

Since the structure of the full ${\cal O}(\alpha^3)$ matrix elements and
the ${\cal O}(\alpha)$ EW corrections in the
pole approximation were discussed in detail in earlier
papers~\cite{Baur:1999kt,Dittmaier:2001ay}, we concentrated on the
phenomenological effects of the non-resonant corrections.
The non-resonant corrections were found to have a very small effect on
the total $W$ cross section and the transverse mass distribution in the 
region $M_T\leq M_W$. However, they increase rapidly in magnitude
with $M_T$ above the $W$ peak, due to the presence of Sudakov-like
electroweak logarithms. Although these corrections are of moderate size
for transverse masses accessible at the Tevatron, they induce a shift of
$\Delta\Gamma_W\approx -7$~MeV in the $W$ width extracted from the tail 
of the transverse mass distribution. Comparison with the expected
overall precision of about $\delta\Gamma_W=25-30$~MeV in Run~II of the
Tevatron shows that it will be
necessary to take the non-resonant electroweak corrections into account
in the data analysis. For transverse masses in the TeV
region which play an important role in new physics searches at the LHC,
the non-resonant electroweak corrections are of the same size as the
${\cal O}(\alpha_s)$ corrections. The strong increase of these
corrections with $M_T$ requires that they are resummed. No such
calculation exists yet for $\ell\nu$ production in hadronic collisions.

We also updated the results of the electroweak $K$-factor for the $W/Z$
cross section ratio $R_{W/Z}$ and the $W$ to $Z$ transverse mass ratio
given in Ref.~\cite{Baur:1999kt} to include the complete ${\cal
O}(\alpha)$ EW radiative corrections and the
${\cal O}(g^4m_t^2/M_W^2)$ corrections to the $Z$ boson cross
section~\cite{Baur:2001ze}.

\acknowledgements
We would like to thank S.~Dittmaier, Y.K.~Kim, A.~Kotwal, K.~McFarland,
M.~Kr\"amer and
W.~Trischuk for useful discussions. We also would like to thank the 
Kavli Institute for Theoretical Physics and the 
Fermilab Theory Group, where part of this work was done, for
their generous hospitality and for financial support.
This research was supported in part by the U.~S.~Department of Energy under
Contract No.~DE-AC02-76CH03000, and the
National Science Foundation under grants, No.~PHY-99-07949,
No.~PHY-0139953 and  No.~PHY-0244875.

\appendix
\boldmath
\section{The purely weak contribution}
\label{sec:weak}
\unboldmath

The pure weak vertex and self energy one-loop corrections to the
process described by the form factor $F_{weak}$ of Eq.~(\ref{eq:full})
can be decomposed into initial and final state contributions as
follows
\begin{equation}\label{eq:weakif}
F_{weak}(\hat s) = F_{weak}^{initial}(\hat s)+F_{weak}^{final}(\hat s)
\end{equation}
with the final state contribution
\begin{equation}
\label{purewf}
F_{weak}^{final}(\hat s) = 
\sum_{j=I,I\!I,I\!I\!I} F_{j,f}^{weak}(\hat s)+\delta Z_1^W-\delta Z_2^W
-\frac{1}{2}\frac{\Sigma_T^{W,weak}(\hat
s)-\Sigma_T^{W,weak}(M_W^2)}{\hat s-M_W^2} 
-\frac{1}{2}\,\delta Z_2^{W,weak}\: .
\end{equation}
Performing the substitution $(f,f')\rightarrow (i,i')$ yields the
corresponding initial state form factor $F_{weak}^{initial}(\hat s)$.
The explicit expressions for the various terms in
Eq.~(\ref{purewf}) can be found in Ref.~\cite{Wackeroth:1996hz}.

\begin{figure}[t]
\begin{center}
\includegraphics[width=12cm]{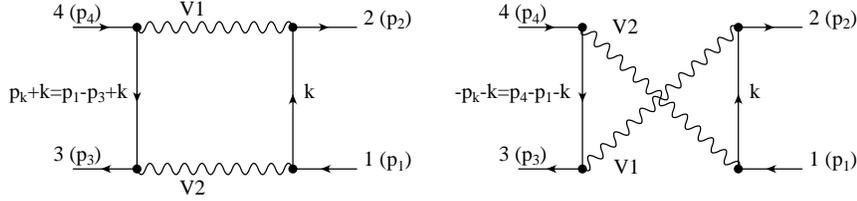}
%\vspace*{2mm}
\caption[]{\label{fig:box} 
Notations used in the calculation of the $V1,V2$ box diagrams.}
\end{center}
\end{figure}

The pure weak box contribution consisting of the $WZ$ box diagrams of
Fig.~\ref{fig:wzbox} in the limit of massless external fermions reads
\begin{eqnarray}
d\hat \sigma_{W\!Z box}(\hat s,\hat t)&=&\frac{\alpha}{\pi}\, d \hat
\sigma^{(0)}(\hat s,\hat t)  
 (\hat s-M_W^2) \sum_{V1,V2=Z,W} 
{\cal R}e (\delta B_{V1V2}+\delta B_{V1V2}^{crossed})
\end{eqnarray}
with 
\begin{eqnarray}
\delta B_{V1V2}(\hat s,\hat t)&=& \lambda_{V1V2} 
\left[2 D_2^0+\hat t (D_1^1+D_1^2+D_1^3+D_2^2+D_2^{23}+D_2^{12})\right]
\nonumber \\
\delta B_{V1V2}^{crossed}(\hat s,\hat u)&=& \lambda_{V1V2}^c
\left [8 D_2^0+\hat u (D_1^1+2 D_1^2+D_1^3+2
(D_2^2+D_2^{23}+D_2^{12}))-2 \hat s D_2^{13})\right] \; , \nonumber \\ 
\end{eqnarray}
where $D_i^{j(k)}=D_i^{j(k)}(\hat s,\hat t,0,M_{V1},0,M_{V2})$ in case
of the box, $\delta B_{V1V2}$, and $D_i^{j(k)}=D_i^{j(k)}(\hat s,\hat
u,0,M_{V1},0,M_{V2})$ in case of the crossed box contribution, $\delta
B_{V1V2}^{crossed}$.  $\lambda_{V1V2}$ and $\lambda^c_{V1V2}$ contain
the dependence on the vector and axial vector parts of the fermion-$Z$
boson couplings which, in the notation of Fig.~\ref{fig:box}, are given
by 
\begin{eqnarray}
\lambda_{W\!Z}&=&2 (v_1+a_1)(v_3+a_3), \nonumber \\
\lambda_{Z\!W}&=&2 (v_2+a_2)(v_4+a_4), \nonumber \\
\lambda_{W\!Z}^c&=&2 (v_1+a_1)(v_4+a_4), \nonumber \\
\lambda_{Z\!W}^c&=&2 (v_2+a_2)(v_3+a_3),
\end{eqnarray}
with $v_i=(I_3^i-2 s_w^2 Q_i)/(2s_wc_w)$ and $a_i=I_3^i/(2s_wc_w)$
parametrizing the $Zf\bar f$ ($f=\ell,\,q$) couplings. Here, $Q_i$ and
$I_3^i$ denote the charge and third component of the weak isospin
quantum numbers of the fermion $i$, and $s_w=\sin\theta_W$,
$c_w=\cos\theta_W$ with $\theta_W$ being the weak mixing angle.

The box integrals arising in the calculation of the Feynman diagrams
of Fig.~\ref{fig:box} are of the form
\begin{eqnarray}\label{eq:int}
\frac{i}{16 \pi^2} (D_0,D^{\mu},D^{\mu \nu})&=& 
\int \frac{d^4 k}{(2\pi)^4 }
\frac{(1,k^{\mu},k^{\mu} k^{\nu})}{[k^2+i \epsilon] 
[(k-p_2)^2-M_{V1}^2] [(k+p_k)^2+i \epsilon] 
[(k+p_1)^2-M_{V2}^2]} \; .\nonumber \\
\end{eqnarray}
The explicit decomposition of the vectorial and tensorial four point
functions 
\begin{eqnarray}
D^{\mu} &=& -p_2^{\mu} D_1^1 + p_k^{\mu} D_1^2 + p_1^{\mu} D_1^3
\nonumber \\[1.mm]
D^{\mu \nu}& =& p_2^{\mu} p_2^{\nu} D_2^1 + p_k^{\mu} p_k^{\nu} D_2^2 +
          p_1^{\mu} p_1^{\nu} D_2^3 + g^{\mu \nu} D_2^0  \nonumber \\[1.mm]
&-&       (p_2^{\mu} p_k^{\nu} + p_k^{\mu} p_2^{\nu}) D_2^{12} -
          (p_2^{\mu} p_1^{\nu} + p_1^{\mu} p_2^{\nu}) D_2^{13} +
(p_k^{\mu} p_1^{\nu} + p_1^{\mu} p_k^{\nu}) D_2^{23} 
\end{eqnarray}
can be found in Ref.~\cite{Denner:1988tv}.

%%%%%%%%%%%%%%%%%%%%%% References %%%%%%%%%%%%%%%%%%%%%%%%%%%%%%%%%%%%%

\bibliographystyle{plain}

\end{document}